\documentclass[preprint,showpacs,preprintnumbers,amsmath,amssymb,floatfix,nofootinbib]{revtex4}

\usepackage{graphicx}

\begin{document}

\preprint{DESY 13-177
\hspace{11.0cm} ISSN 0418-9833}

\title{Inclusive charmed-meson production from bottom hadron decays at the LHC.}

\author{Paolo Bolzoni}
\email{paolo.bolzoni@desy.de}
\author{Gustav Kramer}
\email{gustav.kramer@desy.de}
\affiliation{{II.} Institut f\"ur Theoretische Physik,
Universit\"at Hamburg, Luruper Chaussee 149, 22761 Hamburg, Germany}

\date{\today}

\begin{abstract}

We present predictions for the inclusive productions of the D meson originating from
bottom hadrons at the CERN LHC in the general-mass variable-flavour-number scheme at
next-to-leading order. We present results using two methods to describe the transition
for $b\rightarrow D$: a two-step transition $b\rightarrow B \rightarrow D$, based on the 
$b\rightarrow B$ fragmentation functions and the spectra for $B\rightarrow D$ as measured 
by CLEO and a one-step transition based on the fragmentation functions for $b\rightarrow D$.
The results of both approaches are compared.

\end{abstract}

\maketitle

\section{Introduction}

The study of heavy flavour (charm or bottom) production in proton-proton collisions at the
LHC (Large Hadron Collider) is an important testing ground for perturbative QCD calculations
in a new energy domain.

Charmed hadrons may be produced in $pp$ collisions either directly or as feed-down from the decay
of excited charm resonances. They may also be produced in weak decays of b-hadrons. The first two 
sourses (direct production and feed-down from higher mass resonances) are usually referred to as 
prompt production. Charmed particles from $b$-hadron decays are called secondary charmed 
hadrons or $B$-feed-down charmed hadrons.

Prompt charmed hadron production cross sections were measured in the central rapidity region
$(|\eta|\leq 1)$ in $p\bar{p}$ collisions at the Fermilab Tevatron collider 
at $\sqrt{s}=1.96\, \rm{TeV}$ \cite{Acosta:2003ax} and in the central rapidity region 
$(|\eta|\leq 0.5)$ in $pp$ collisions at $\sqrt{s}=2.76\, \rm{TeV}$ \cite{Abelev:2012vra} and
at  $\sqrt{s}=7\, \rm{TeV}$ \cite{Abelev:2012tca,ALICE:2011aa} and in the forward rapidity 
region $(2.0\le y \le 4.5)$ \cite{Aaij:2013mga} at the CERN LHC. In addition, charmed hadron 
production cross section measurements in which prompt production and $B$-feed-down production
have not been separated, were reported by the ATLAS Collaboration 
\cite{ATLAS:2011wsp,ATLAS:2011fea} at the LHC. Perturbative calculations of these charmed 
hadron production cross section at next-to-leading order based on the General Mass Variable
Flavour Number Scheme (GM-VFNS) \cite{Kniehl:2012ti} and on the fixed order with 
next-to-leading-logarithmic resummation (FONLL) \cite{Cacciari:2012ny} reproduce the measured
cross sections \cite{Abelev:2012vra,Abelev:2012tca,ALICE:2011aa,Aaij:2013mga,ATLAS:2011wsp}.

It is conceivable that in the near future prompt charmed hadron production and production from
weak decays of $B$ meson could be well separated and that the production cross section for
$B$-feed-down production could be measured. Actually in the case of inclusive lepton production 
from heavy hadron decays such a separation has been achieved. In two experiments the inclusive 
leptonic production cross section is measured separately for solely $b$-hadron decays 
\cite{Khachatryan:2011hf,Abelev:2012sca}. In particular, in the ALICE measurement 
\cite{Abelev:2012sca} the production cross section of electrons from semileptonic bottom hadron
decays was selected by using the information on the distance of the secondary decay vertex 
displaced in space from the primary collision vertex. It might be possible to apply this 
technique also in other decays of bottom hadrons, as for example, $B\rightarrow D+X$, where
$D$ is any charmed meson, $D^0$, $D^\pm$, $D^{*\pm}$, $D^{*0}$ or $D_s$.
The calculation of the charm hadron production from inclusive $B$ decay into these charm hadron
can be done in two ways. Either one uses fragmentation functions (FFs) for the fragmentation
process $b\rightarrow D+X$, where $D$ is one of the charmed mesons. Such FFs have been constructed
in the past from data on $e^+e^-\rightarrow D+X$ and will be described in more details in the 
next section. The other way is to calculate the inclusive cross section $p+p\rightarrow B+X$, 
where $B$ is anyone of the bottom mesons $B^0$, $B^+$ and $B_s$ or its antiparticles and then 
convolute this cross section with the inclusive spectrum of $B\rightarrow D+X$, which is 
known from measurements of the CLEO collaboration at CESR, the 
Cornell Electron Storage Ring \cite{Gibbons:1997ag}, similar as we did for the inclusive 
production of leptons from semileptonic $B$ decays \cite{Bolzoni:2012kx}. 
For this calculation one needs the FFs for $b\rightarrow B$, which are knwon from the 
work in \cite{Kniehl:2008zza} and which we shall use. Instead of using the experimental information on the inclusive spectrum of $B \to D+X$ one could think of
calculating this spectrum and compare it with the CLEO data 
\cite{Gibbons:1997ag}. First attempts in this direction have been done quite some time ago by Wirbel
and Wu \cite{Wirbel:1989ww} much earlier than the CLEO data had appeared.
Unfortunately this comparison of models as the one proposed in \cite{Wirbel:1989ww} to the CLEO data has not been done so far. Therefore we shall rely in our calculations on the empirical D meson spectrum from B decay as measured in 
\cite{Gibbons:1997ag}. After convolution with the $b \to B$ FFs one obtains the FFs for $b \to D$. This is an alternative method for calculationg these FFs, which is essentally based on the $b \to B$ FFs from $e^+e^-$ annihilation data obtained at
LEP and SLC. Of course it depends on the tagging of charmed mesons by CLEO at the
$\Upsilon (4S)$ resonance as opposed to the direct measurements of these mesons
performed by the LEP experiments at the $Z$ resonance to be described in the next section.. The main difference of the two approaches is the tagging of the $D$
mesons at the two energies, the CLEO energy at $\sqrt{s}=10.58$ GeV versus the
LEP energy $\sqrt{s}=m_Z$.

The content of this paper is as follows. In Section \ref{two} we describe the input choices 
of parton distribution functions (PDFs) and $B$- and $D$-meson FFs. In this section we also explain 
how the fragmentation of $B$ into $D$ mesons has been obtained from the CLEO data. In Section \ref{three} we present 
our predictions of the GM-VFN scheme for the cross sections $p+p\rightarrow B+X\rightarrow D+X^{'}$ for
the four cases: $D^0$, $D^\pm$, $D^{*\pm}$ and $D^{*0}$. In addition we give the corresponding cross
sections where the fragmentation functions $b\rightarrow D^0$, $D^\pm$ and $D^{*\pm}$ have been used in terms
of ratios to the cross sections for prompt production based on the FFs for $c\rightarrow D^0$, $D^\pm$
and $D^{*\pm}$ and compare the ratios for the two approaches of calculating the inclusive $D$ meson 
production cross sections from bottom quarks.

\section{Input PDFs, FFs and Setup}
\label{two}

The calculations presented in this paper are performed in the theoretical framework of the
GM-VFNS approach for $pp$ collisions which has been presented in detail in Refs.
\cite{Kniehl:2008zza,Kniehl:2004fy,Kniehl:2005mk}. In this Section we describe our choice 
of input for the calculation of inclusive production of various $D$ meson species originating
from bottom quarks. For the ingoing protons we use the PDF set CTEQ$6.6$ \cite{Nadolsky:2008zw}
as implemented in the LHAPDF library \cite{Whalley:2005nh}. This PDF set was obtained in the 
general-mass scheme using the input mass values $m_c=1.3\,\rm{GeV}$, $m_b=4.5\,\rm{GeV}$, and
for the QCD strong coupling $\alpha_s^{(5)}(m_Z)=0.118$. The $c$- and the $b$-quark PDFs 
have the starting scale $\mu_0=m_c$ and $\mu_0=m_b$, respectively.

The nonperturbative FFs for the transition $b\rightarrow B$ needed for the approach where $D$
production is calculated from $B$ decay to $D$ mesons, were obtained by a fit to $e^+e^-$ 
annihilation data from the ALEPH \cite{Heister:2001jg}, OPAL \cite{Abbiendi:2002vt} and
SLD \cite{Abe:1999ki,Abe:2002iq} collaborations and have been published in \cite{Kniehl:2008zza}.
The combined fit to the three data sets was done using the NLO sclae parameter
$\Lambda^{(5)}_{\overline{MS}}=227\,\rm{MeV}$ which corresponds to $\alpha_s^{(5)}(m_Z)=0.1181$
adopted from \cite{Nadolsky:2008zw}. Consistent with the chosen PDF, the starting scale of the 
$b\rightarrow B$ FF was assumed to be $\mu_0=m_b$, while the $q,g\rightarrow B$ FFs, where $q$
denotes the light quarks including the charm quark, are assumed to vanish at $\mu_0$. Indeed 
their contribution is very small since they appear only via the evolution of the FFs to 
larger scales. As input we used the FFs with a simple power Ansatz which gave the best fit to
the experimental data. The bottom mass in the hard scattering cross sections is $m_b=4.5 \,\rm{GeV}$
as it is used in the PDF CTEQ$6.6$ and in the FFs for $b\rightarrow B$.

For comparison with the results for the prompt production and for calculating the ratios of
the various contributions, we also need the FFs for the transitions $c\rightarrow D^0$, $D^+$ and
$D^{*+}$ which we take from \cite{Kneesch:2007ey}. There we used the so-called Global-GM fit
which includes fitting in addition to the OPAL data \cite{Alexander:1996wy} together with the most 
precise data on $D$ meson production from the CLEO Collaboration at CESR \cite{Artuso:2004pj}
and from the Belle Collaboration at KEKB \cite{Seuster:2005tr}. The fits in \cite{Kneesch:2007ey},
which by including the OPAL data from LEP$1$, yield also the FFs for $b\rightarrow D$.They are based 
on the charm mass $m_c=1.5\,\rm{GeV}$, which is slightly larger than the one used in the 
CTEQ$6.6$ PDFs. The starting scale for $c\rightarrow D$ is $\mu_0=m_c$, as it is for the 
$g,q\rightarrow D$ FFs, whereas for the $b\rightarrow D$ FF it is $\mu_0=m_b$. The FFs for 
$b\rightarrow D$ as given in \cite{Kneesch:2007ey} are used for the second approach for the 
B feed-down production in $pp$ collisions at the LHC.

The theoretical accuracy of the theoretical prediction is estimated by calculating the cross 
sections with varying renormalization and factorization scales $\mu_{\rm{R}}$, $\mu_{\rm{I}}$ 
and $\mu_{\rm{F}}$, denoting the renormalization and the factorization scales of initial and 
final state singularities respectively. We choose the scales to be of order $m_T$, where $m_T$
is the transverse mass $m_T=\sqrt{p_T^2+m^2}$ with $m=m_b$ for the case of the bottom quark 
and $m=m_c$ for charm quark production. For exploiting the freedom in the choice of scales we have
introduced the scale parameters $\xi_i$ ($i=\rm{R,I,F}$) by $\mu_i=\xi_i m_T$. We vary as usual 
the values of the $\xi_i$'s independently by a factor of two up and down while keeping any
ratio of the $\xi_i$ parameters smaller than or equal to two. The uncertainties due to the scale 
variation are dominant. Therefore PDF related uncertainties and variations of the bottom and
charm mass are not considered.

The fragmentation of the final state partons $i$ into $D$ mesons ($D=D^0$,$D^+$,$D^{*+}$ 
and $D^{*0}$) is calculated from the convolution
\begin{equation}
\label{convol}
D_{i\rightarrow D}(x,\mu_F)=\int_x^1\frac{dz}{z}\, 
D_{i\rightarrow B}\left(\frac{x}{z},\mu_F\right) \frac{1}{\Gamma_B}\frac{d \Gamma}{dz}(z,P_B).
\end{equation}
In this formula, which is quite analogous to the formula we used for the fragmentation of 
partons $i$ into leptons \cite{Bolzoni:2012kx}, $D_{i\rightarrow B}(x,\mu_F)$ is the
nonperturbative FF determined in \cite{Kniehl:2008zza} for the transition $i\rightarrow B$, 
$\Gamma_B$ is the total $B$ decay width and finally $d\Gamma(z,P_B)/dz$ is the decay spectrum
of $B\rightarrow D$. For a given $D$ meson transverse momentum $p_T$ and rapidity $y$, $P_B$
is given by $P_B=|\vec{P_B}|=\sqrt{p_T^2+m_T^2\sinh^2 y}/z$. The decay distribution
$d\Gamma/dk'_L$, where the momentum $k'_L$ is parallel to $\vec{P_B}$ is obtained from the decay 
distribution in the rest system of the $B$ meson using the formula in Eq.(3.16) in 
Ref.\cite{Kniehl:1999vf}, where the formula was derived for the decay $B\rightarrow J/\Psi+X$
instead of $B\rightarrow D+X$. From this one obtaines $d\Gamma(z,P_B)/dz$ used in Eq.(\ref{convol})
with $z=k'_L/P_B$.

\begin{figure*}
\includegraphics[width=7.5cm]{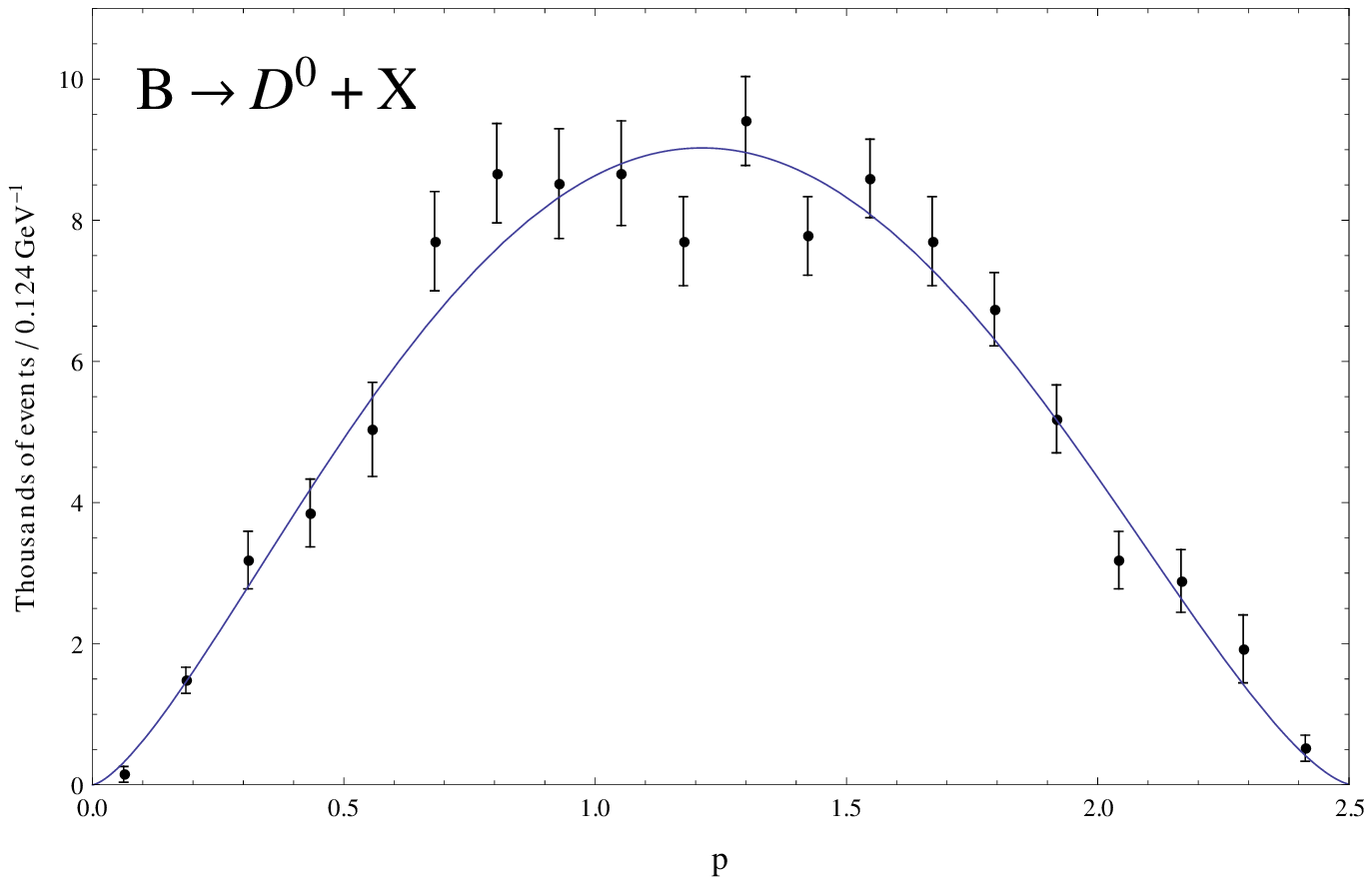}
\includegraphics[width=7.5cm]{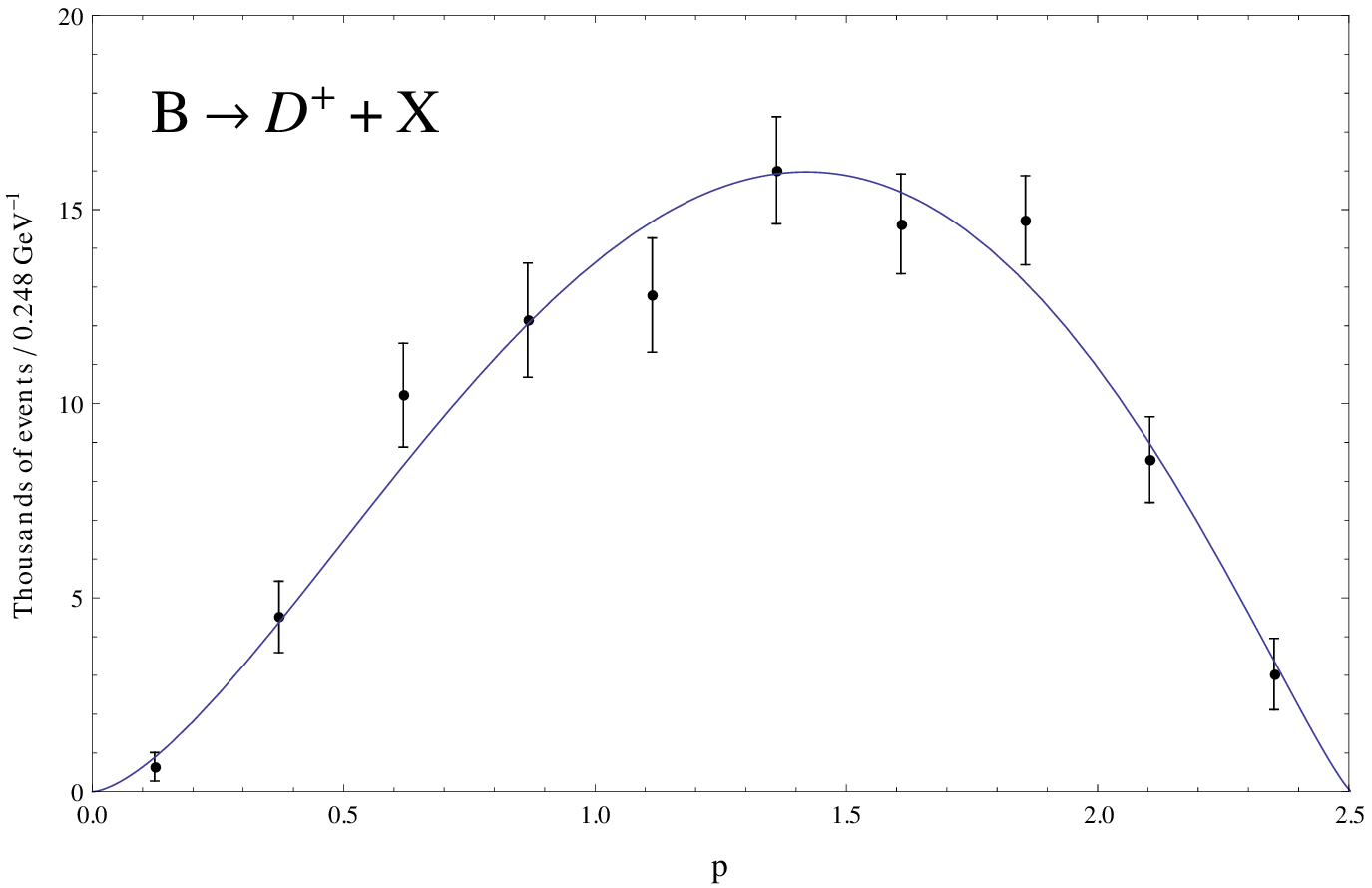}
\includegraphics[width=7.5cm]{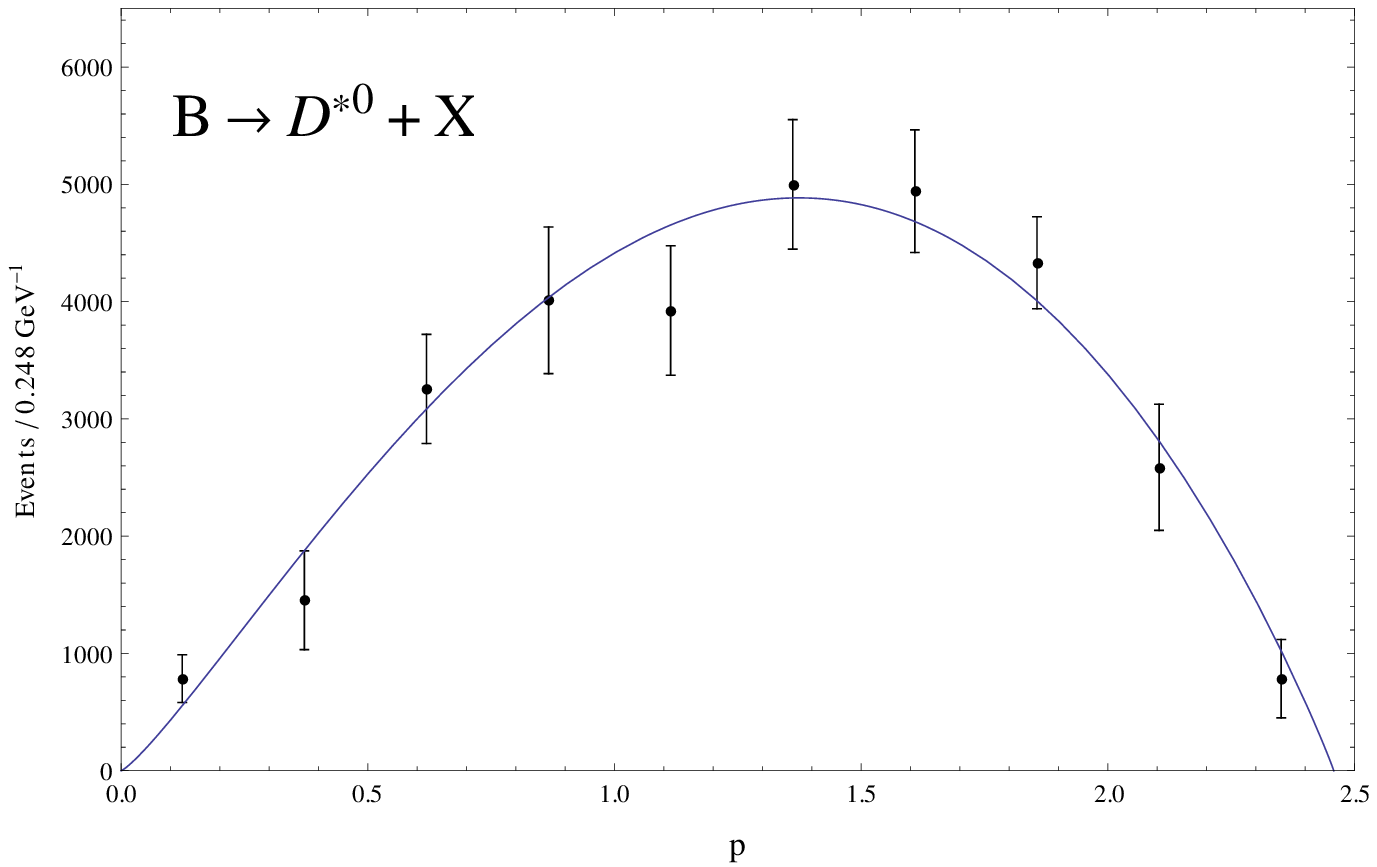}
\includegraphics[width=7.5cm]{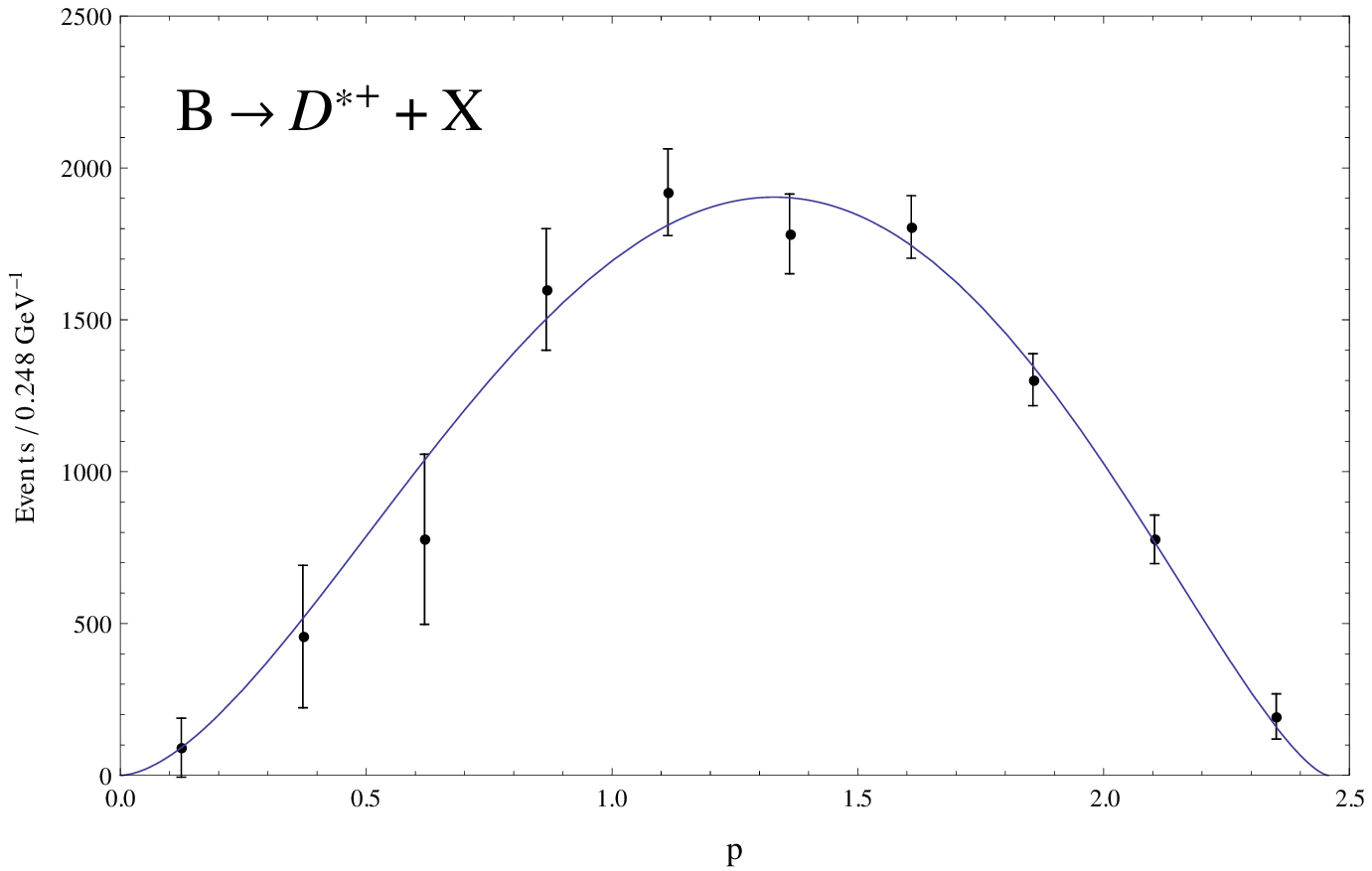}
\caption{\label{fig:fits} Fits of data for the spectrum of the 
$B$ meson decays into $D=D^0$, $D^+$, $D^{*0}$ and $D^{*+}$.}
\end{figure*}

The momentum ($p$) spectra in inclusive decays $B\rightarrow D+X$ ($D=D^0,D^+,D^{*0},D^{*+}$) have been
measured as a function of $x=p/p_{\rm{max}}$ in \cite{Gibbons:1997ag} and they are given in graphical
form in Figs. $16,21,25$ and $33$ of this reference. The data points have been read off from these
figures and fitted by a simple power law in $p$ of the form
\begin{equation}\label{fit}
f(p)=N\, p^\alpha\, (a-p)^\beta.
\end{equation} 
The function $f(p)$ is related to the partial $D$ decay spectrum according to $d\Gamma/dp=c f(p)$.
\begin{table}
\centering
\begin{tabular}{|c|c|c|c|c|c|c|c|}
\hline
Channel & $p_{\max}[\rm{GeV}]$ & $N$ & $\alpha$ & $\beta$ & $a$ & $c$ & $\chi^2_{\rm{dof}}$\\
\hline
$B\rightarrow D^0X$ & 2.5070 & 4.5603 & 1.4502 & 1.5560 & 2.5070 & 0.047139 & 1.15 \\
$B\rightarrow D^+X$ & 2.5050 & 8.3427 & 1.5727 & 1.2013 & 2.5050 & 0.00098725 & 0.66 \\
$B\rightarrow D^{*0}X$ & 2.4578 & 3081.4 & 1.2084 & 0.9538 & 2.4578 & 3.3384$\cdot 10^{-5}$ & 0.58\\
$B\rightarrow D^{*+}X$ & 2.4568 & 976.3 & 1.7290 & 1.4648 & 2.4568 &  8.3617$\cdot 10^{-5}$ & 0.34\\
\hline
\end{tabular}
\caption{The fitted parameters $c$, $N$, $\alpha$, $\beta$ and $p_{\rm{max}}$ for the various
channels together with the corresponding $\chi^2$ per degree of freedom.}\label{parameters}
\end{table}
The obtained parameters $c$, $N$, $\alpha$, $\beta$ and $p_{\rm{max}}$ are collected in Table 
\ref{parameters} for the four cases $D=D^0$, $D^+$, $D^{*0}$ and $D^{*+}$. In the last column 
we report also the corresponding $\chi^2$ per degree of freedom.
Integrating the fits $d\Gamma/dp$ over $p$ in the kinematic range $0\leq p\leq p_{\rm{max}}$
yields the branching ratios for $B\rightarrow DX$. The result for 
$D=D^0$, $D^+$, $D^{*0}$ and $D^{*+}$ is 0.627, 0.237, 0.260 and 0.225, respectively\footnote{These branching ratios are taken from the recent PDG 
\cite{Beringer:1900zz} values. They are given for the decay of a mixture of 
$B^+$ and $B^0$ mesons.}.
The corresponding numbers reported by the CLEO collaboration are $0.636\pm 0.030$, 
$0.235\pm 0.027$, $0.247\pm 0.028$ and $0.239\pm 0.020$, where the statistical and sytematic
errors have been combined in quadrature, are in satisfactory agreement with the 
values obtained in our fits. In the CLEO experiment the inclusive $D$ decays arise from a 
mixture of $B^0$ and $B^+$. The quality of our fits can be seen from Fig. \ref{fig:fits} 
where the CLEO data together with our fits are shown for $B\rightarrow D^0$, 
$D^+$, $D^{*0}$ and $D^{*+}$ as a function of p. 
Actually, the inclusive $D$ meson spectra as measured by CLEO are not measured in the $B$
rest system, but produced at the $\Upsilon(4S)$ resonance mass. Since this mass (10.58 GeV)
is slightly above the threshold for $B$ meson pair production (10.56 GeV) the $B$ mesons 
are not at rest and have a momentum ranging from 265 MeV up to 355 MeV. This motion smears 
the value of $p$ or $x$ relative to what it would be if the $B$ were at rest. We studied this
effect in our earlier work on the inclusive lepton spectra in $B$ decays \cite{Bolzoni:2012kx}
and we found a very negligible effect on the spectra. Therefore we did not consider this effect
for the inclusive $D$ meson spectra, \emph{i.e.} we consider our fits in Eq.(\ref{fit}) 
of the CLEO data as the spectra produced for $B$ mesons at rest. With the parametrization 
of the $D$ meson spectra known in the $B$ rest system we calculated the $D$ meson spectra
in the moving system as a function of $k'_L=x P_B$ where $k'_L$ is the $D$ meson momentum
that is parallel to $\vec{P_B}$ using Eq.(3.16) in \cite{Kniehl:1999vf} with $M_{\psi}$ replaced 
by the rest mass of the respective $D$ meson.

The exact formula for $d\Gamma/dx$ as given in \cite{Kniehl:1999vf} is rather difficult to
evaluate since $P_B$ depends on $p_T$ and the rapidity $y$ of the produced $D$ meson.
Therefore, as in our previous work \cite{Bolzoni:2012kx} we applied for $d\Gamma/dx$
the asymptotic formula, also given in \cite{Kniehl:1999vf}. This is valid for $P_B\gg M_B$,
where $M_B$ is the mass of the $B$ meson. We calculated $d\Gamma/dx$ for various $P_B$ and found
that the exact formula differs from the asymptotic formula by less than $5\%$ for $P_B=15\,\rm{GeV}$.
This is approximately achieved for $p_T=5\,\rm{GeV}$ and $y=0$ since $d\Gamma/dx$ is 
peaked at small $x\simeq 0.3$.

The alternative approach, which we used to calculate $d\sigma/dp_T$ for the production
of $D$ mesons originating from feed-down $B$ meson production is the calculation 
with FFs for $b\rightarrow D$. Such FFs for $D^0$, $D^+$ and $D^{*+}$ are available in 
\cite{Kneesch:2007ey} from fits to CLEO, Belle, OPAL and ALEPH $e^+e^-$ annihilation
data. In order to distinguish contributions originating from $c\rightarrow D$ and
$b\rightarrow D$ one needs data, where the $b\rightarrow D$ contributions are separated
from the total $e^+e^-$ annihilation cross sections into charmed hadrons. This has 
been achieved at LEP1 at the Z resonance by the OPAL \cite{Alexander:1996wy,Ackerstaff:1997ki}
and the ALEPH \cite{Barate:1999bg} collaborations.
Apart from the full cross sections, they also determined the contribution from $Z\rightarrow b\bar{b}$
decays. Using these data we have determined the FFs for $c\rightarrow X_c$ and $b \rightarrow X_c$
for charmed hadrons $X_c=D^0$, $D^+$, $D^+_s$ and $\Lambda^+_c$ already in 
\cite{Kniehl:2005de,Kniehl:2006mw} and earlier references therein. The FFs constructed in \cite{Kneesch:2007ey}
have the advantage that they include the data from CLEO \cite{Artuso:2004pj} and Belle 
\cite{Seuster:2005tr} and in this way the FFs for $c\rightarrow X_c$ are much better constrained
than in the earlier works \cite{Kniehl:2005de,Kniehl:2006mw}.

\section{Results and comparison of the two approaches}
\label{three}

In this Section we collect our results for the cross sections $d\sigma/dp_T$ as a function of
$p_T$ for the two approaches described in the previous Section. For the rapidity range 
we choose $-0.5\leq y \leq 0.5$ as used in the ALICE experiment \cite{Abelev:2012vra},
over which the cross section $d^2\sigma/(dp_T dy)$ is integrated over.
The basic formalism needed in the first approach is based on the cross section $d\sigma/dp_T$
for the inclusive production $pp\rightarrow BX$ and is described in detail in the work 
\cite{Kniehl:2011bk} and the references given there. In this work the inclusive $B$ meson
production cross section in $p\bar{p}$ collisions at $\sqrt{s}=1.96~\rm{TeV}$ and in $pp$
collisions was calculated and good agreement with the respective data from the CDF run II 
\cite{Acosta:2004yw,Aaltonen:2009xn,Abulencia:2006ps} and also with the data from the CMS
Collaboration \cite{Khachatryan:2011mk,Chatrchyan:2011pw,Chatrchyan:2012xg} at the LHC at
$\sqrt{s}=7~\rm{TeV}$ was found. 
\begin{figure*}
\includegraphics[width=7.5cm]{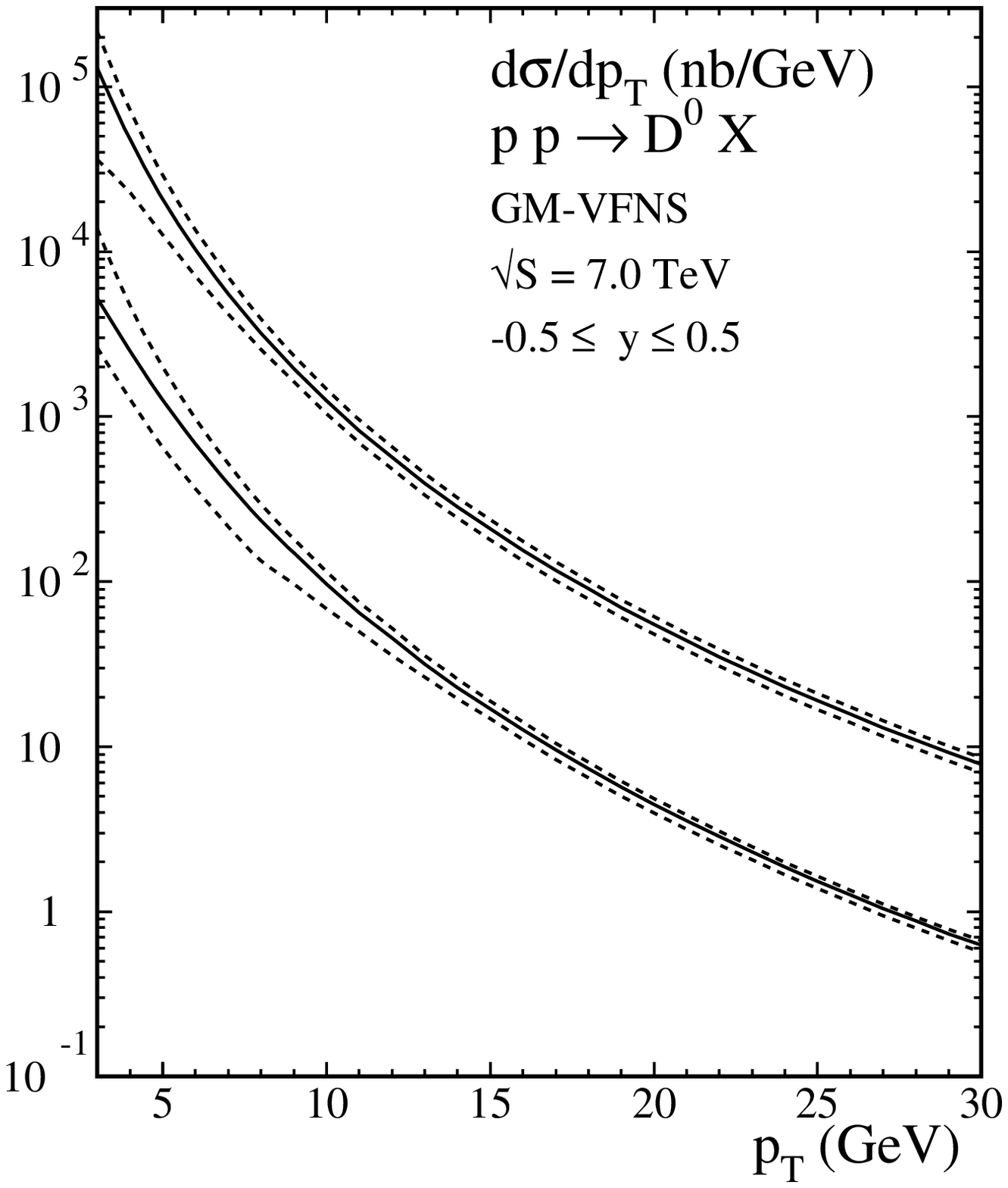}
\includegraphics[width=7.5cm]{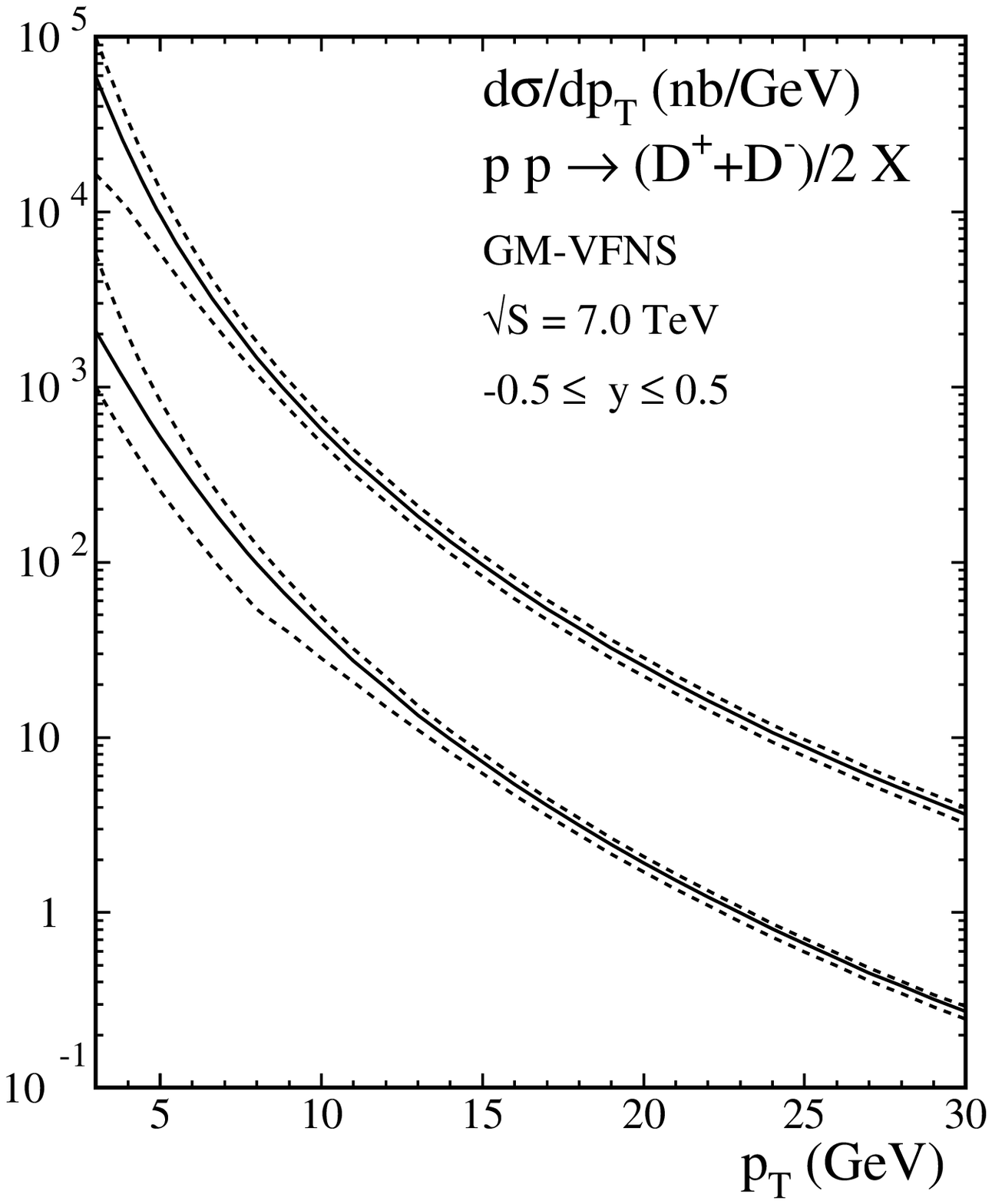}
\includegraphics[width=7.5cm]{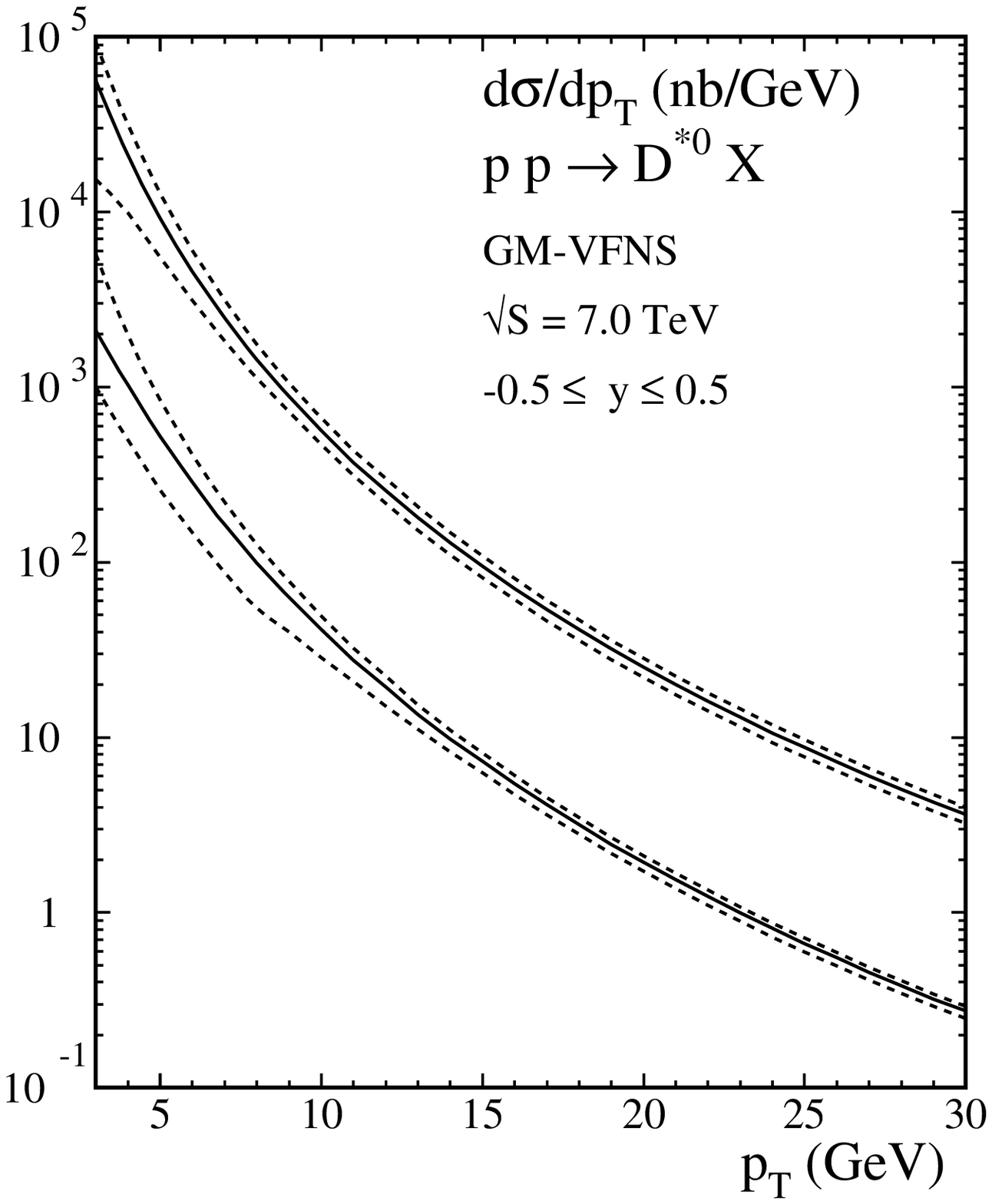}
\includegraphics[width=7.5cm]{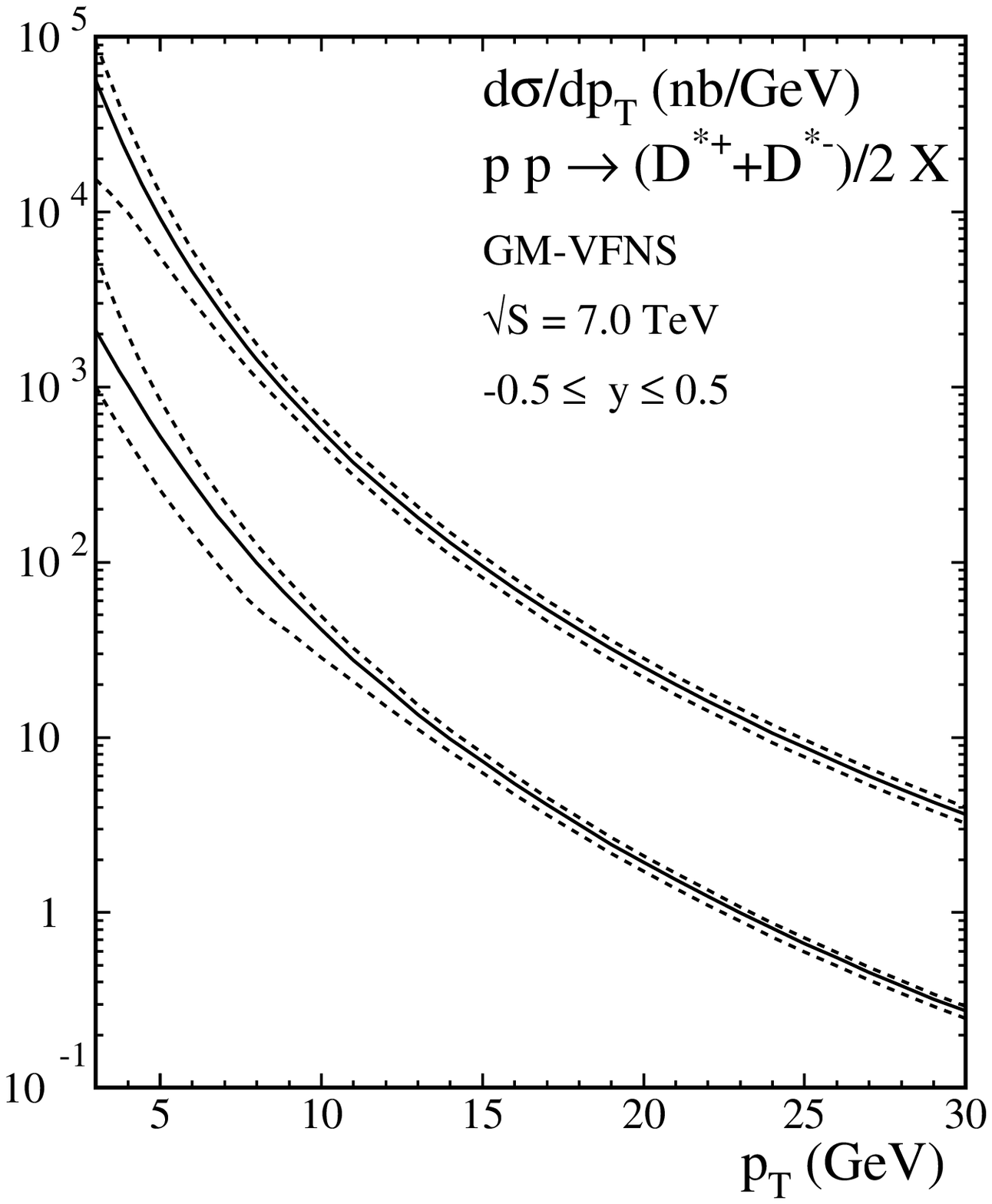}
\caption{\label{fig:main} The lower curves, together with the corresponding scale uncertainties
(dashed lines), are our predictions for the $B$-feed-down hadrons 
($b\rightarrow B \rightarrow D$). For comparison we also show (upper curves) the cross sections corresponding to
$c\rightarrow D$.}
\end{figure*}
\begin{figure*}
\includegraphics[width=7.5cm]{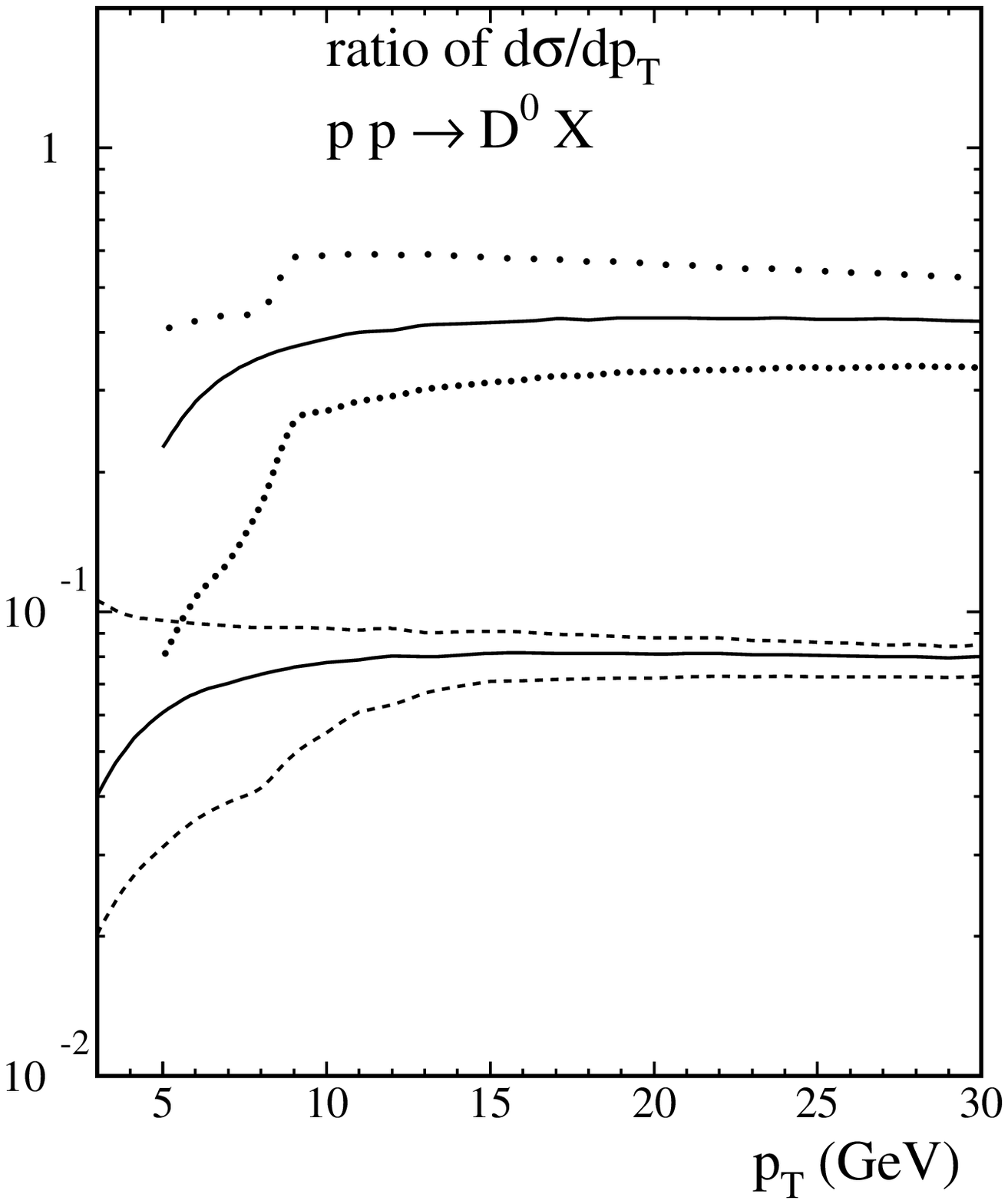}
\includegraphics[width=7.5cm]{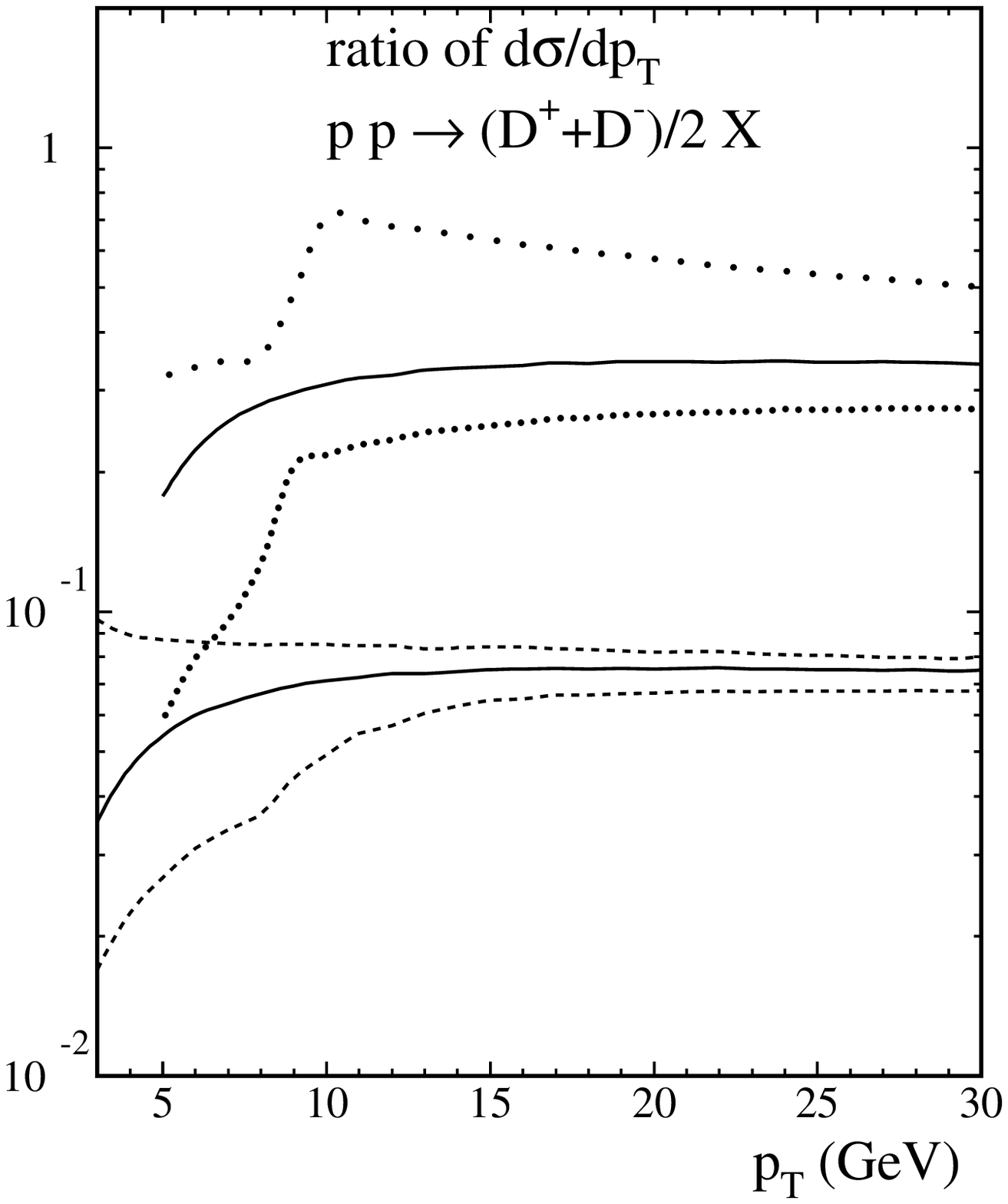}
\includegraphics[width=7.5cm]{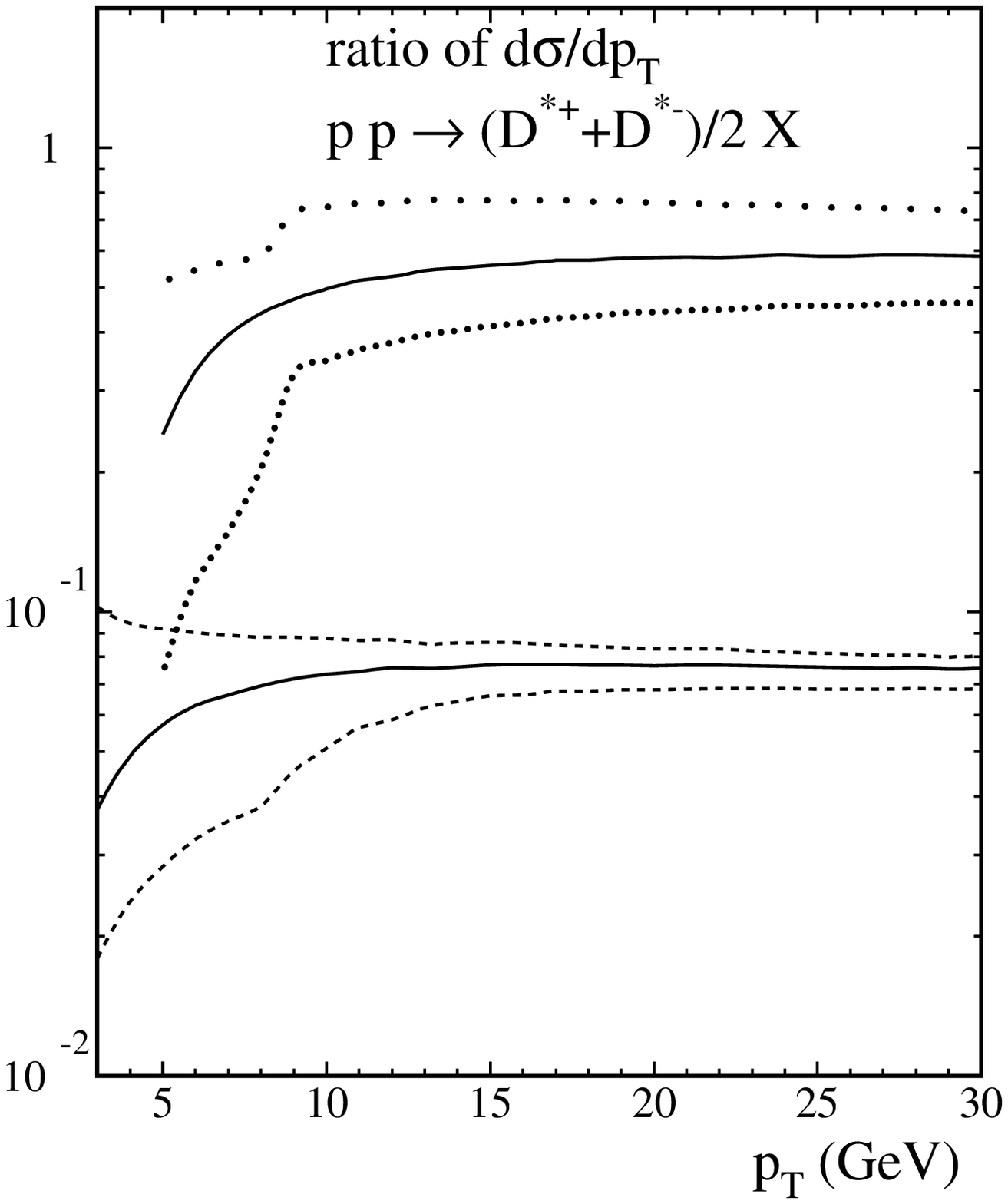}
\caption{\label{fig:ratio} Ratio of the cross section for $b\rightarrow B \rightarrow D$ (lower curves)
and for $b\rightarrow D$ (upper curves) to the
cross section for $c\rightarrow D$.   
The last ratios are multiplied by a factor of 10 for better visibility.}
\end{figure*}

Our main results are shown in Fig. \ref{fig:main} (lower curves) where we present $d\sigma/dp_T$
integrated over $-0.5\leq y \leq 0.5$ as a function of $p_T$ for $3\leq p_T \leq 30\,\rm{GeV}$ for
$p+p\rightarrow D^0X$, $(D^++D^-)/2\,X$, $(D^{*+}+D^{*-})/2\,X$ and $D^{*0}\,X$ for the default
scale $\xi_{R}=\xi_I=\xi_F=1$ (solid line) and for scales which lead to maximal and minimal cross sections
(dashed lines) inside the constraints for $\xi_R$, $\xi_I$ and $\xi_F$ defined in the previous Section. 
The cross section for the $D^0+X$ final state is larger than the other three due to the larger branching 
fractions for $B^+/B^0\rightarrow D^0/\bar{D^0}\,X$ than, for example, for 
$B^+/B^0\rightarrow D^\pm\,X$. The cross sections $d\sigma/dp_T$ for the final states $D^{*+}X$ and
$D^{*0}X$ are also slightly different due to the different branching fractions for 
$B\rightarrow D^{*+}\,X$ and $B\rightarrow D^{*0}\,X$ and differences in the spectral shapes as can be
seen in Fig. \ref{fig:fits}. Actually the results for $D^0$ and $D^{*0}$ are alo for the averages of $D^0$ and $\overline{D^0}$ and $D^{*0}$ and $\overline{D^{*0}}$, respectively.

In Fig. \ref{fig:main} we have plotted also the cross sections for $d\sigma/dp_T$ based on 
$c\rightarrow D^0\,X$, $D^{\pm}\,X$, $D^{*\pm}\,X$ and $D^{*0}\,X$, just for comparison with the 
corresponding feed-down $B$ cross sections. These cross sections are the same as those
presented in \cite{Kniehl:2012ti}, which have been compared there to the ALICE experimental
data \cite{ALICE:2011aa}. The shapes of the two cross sections for $c\rightarrow D$
and $b\rightarrow B\rightarrow D$ are very similar, at least for the larger $p\geq 10\,\rm{GeV}$.
This is seen more clearly in Fig. \ref{fig:ratio} where we have plotted the ratio of $d\sigma/dp_T$ for the two 
cross sections (lower curves), for the default scale choice (full curve) and the maximal and 
minimal scale choice (dashed curves). 
For the latter two curves the scale variation is included only for the numerator, \emph{i.e.} 
for $d\sigma/dp_T$ for $b\rightarrow B\rightarrow D$. As it can be seen, this ratio is nearly 
independent of $p_T$ for approximately $p_T>10\,\rm{GeV}$ and is around $8.0\%$, $7.5\%$ and $7.5\%$ for the largest $p_T$ and for the three cases $D^0$ ,$D^{\pm}$ and $D^{*\pm}$, respectively.
For $p_T<10\,\rm{GeV}$ this ratio decreases and apporaches $4\%$ at $p_T=3\,\rm{GeV}$.

The predictions for $d\sigma/dp_T$ using the one-step approach based on the fragmentation functions for $b\rightarrow D$
from \cite{Kneesch:2007ey} are given only in the form of ratios to $d\sigma/dp_T$ for $c\rightarrow D$.
These results are also shown in Fig. \ref{fig:ratio} as the upper curves: full for the default 
and dotted curves for the maximal and minimal result. For better visibility these ratios are multiplied
by a factor 10. The shape of this ratio as a function of $p_T$ for $5.0<p_T < 30\,\rm{GeV}$
is similar as for the case of the two-step FFs: $b\rightarrow B\rightarrow D$. The only significant 
differences are the absolute values of the ratios. At the largest $p_T$ the ratios for $b \rightarrow D$
are approximately between $50\%$ and $20\%$ smaller than the ratios for $b \rightarrow B \rightarrow D$ shown
as the lower curves in Fig. \ref{fig:ratio}. The reason for this decrease is not caused by a mismatch of the branching ratios. Indeed, for example, in the case 
$b \rightarrow B \rightarrow D^0$ the total branching ratio is approximatel $0.8\times 0.627=0.502$
in good agreement with the branching ratio for $b\rightarrow D^0$ equal to $0.515$
as given in Table 11 of \cite{Kneesch:2007ey} and similarly for the other channels. Therfore the reason for the differeent ratio in the upper and lower part of
Fig. \ref{fig:ratio} must lie in the shape of the FFs. The FFs in the two approaches have a similar shape of approximately gaussian form with a maximum near
$z=0.25$.   

First we should ask, however, why the cross section $d\sigma/dp_T$ is so much smaller for $b\rightarrow B \rightarrow D$ than the one for
$c \rightarrow D$ as already clearly seen in Fig. \ref{fig:main}. To answer this
question we calculated the average $z$ as a function of $p_T$ in the relevant region of $p_T$. This $<z>(p_T)$ is the quantity
\begin{equation}
  <z>(p_T) = \frac{\int dzzd\sigma/dp_T}{\int dzd\sigma/dp_T}
\end{equation}
where $z$ is the scaling variable of the respective  FFs. It is understood that the integration of the quantities of the numerator and of the denominator is done over the the rapidity interval $|y| \leq 0.5$. The range of $<z>$'s for the four cases $D^0, D^{\pm}, D^{*\pm}$ and $D^{*0}$ is in the $p_T$ range $3.0 \leq
p_T \leq 30$ GeV equal to $<z> = [0.43,0.48],[0.45,0.50],[0.35,0.45]$ and $[0.45,0.50]$. This means that the relevant $z$ range, where the FFs contribute to
$d\sigma/dp_T$, lie outside the maximum of the FFs on the right side, where the FFs decreased already by an aprecciable factor. The FFs for $c \rightarrow D$, on the other hand, have a similar shape with a maximum at $<z> \cong 0.65$
\cite{Kneesch:2007ey}, which is approximately the relevant range for $d\sigma/dp_T$. For example, for $c \rightarrow D^0$ we obtain $<z>=[0.67,0.64]$ in the range $3<p_T<30$ GeV, which is just in the vicinity of the $z$, where the corresponding FF is maximal. From this we conclude that the smallness of the cross section $d\sigma/dp_T$ for $b \rightarrow D$'s in the considered $p_T$ range is due to the fact that the corresponding FFs contribute only outside the range where the
FF is maximal and has decreased already appreciably.

For the one-step FFs $b \rightarrow D$'s the average $z$ ranges in the cross section $d\sigma/dp_T$ are similar as in the two-step
process. The only exception is $b \rightarrow D^{*\pm}$ where $<z>$ ranges in the
interval $[0.44,0.53]$, which is $0.1$ larger than for the case $b \rightarrow B\rightarrow D^{*\pm}$., where the corresponding FF is smaller, which has the effect that the cross section $d\sigma/dp_T$ is snaller by $20\%$ than for the two- step process as can be seen in Fig. \ref{fig:ratio}. In the other two cases $D^0$
and $D^{\pm}$ it turns out that in the relevant region the one-step FF is smaller than the two-step FF.
\section{Summary}

In this paper we have calculated the cross sections $d\sigma/dp_T$ for inclusive
$D$ meson production for several $D$ meson charge states originating from weak bottom quark decays at the LHC c.m. energy of $7$ TeV in the framework of the
GM-VFN scheme. We used two approches for calculating the FFs for $b \rightarrow 
D$. In the two-step approach: $b \rightarrow B \rightarrow D$'s we constructed 
the FFs based on the $b \rightarrow B$ FF fitted to $B$ production data from
LEP and SLC convoluted with FFs for $B \rightarrow D$'s obtained from measuerements of the CLEO collaboration. In the one-step approach we used FFs for
$b \rightarrow D$, constructed from $e^+e^-$ annihilation cross section for the production of $D$ mesons from $b$ quarks.
 
In both approaches the cross section for $pp \rightarrow DX$ with $D$'s originating from $b$ quarks decays is only a small fraction of the dominant contribution where the $D$ mesons come from the fragmentation of charm quarks. The reason for the reduction of the $pp \rightarrow bX' \rightarrow DX$ cross sections as compared to the $pp \rightarrow cX' \rightarrow DX$ originates from the fact that in the second case the FF for $c \rightarrow D$'s contributes in the region of fractional momenta, where the FF is maximal, whereas in the first case the contributing region of the FF is above the maximal region where the FF has decreased aprecciably. The difference in the cross sections for the one-step and two-step approaches
can be explained by the stronger fall-off of the FFs from the one-step approach above the $z$'s where the maximum occurs. Therfore the measurement of the cross section for $D$ meson production from $b$ decays at the LHC is an ideal place to get information on the behaviour of the FFs for $b \rightarrow D$'s beyond where the maximun occurs.

\section*{Acknowledgments}

We thank H. Spiesberger for help with the modification of the GM-VFNS cross 
section routine.  
This work was supported in part by the German Federal Ministry
for Education and Research BMBF through Grant No.\ 05~H12GUE, 
by the German Research Foundation DFG through Grant No.\ 
KN~365/7--1, and by the Helmholtz Association HGF through 
Grant No.\ Ha~101.
%
%

\end{document}